\documentclass[aps,prl,twocolumn, showpacs]{revtex4}

\usepackage{epsfig}
\usepackage{graphicx}

\def\be{\begin{equation}}
\def\ee{\end{equation}}
\def\bea{\begin{eqnarray}}
\def\eea{\end{eqnarray}}

\begin{document}

\title{Graphene with Structure-Induced Spin-Orbit Coupling:\\
Spin-Polarized States, Spin Zero Modes, and Quantum Hall Effect}
\author{Emmanuel I. Rashba}
\affiliation{Department of Physics and Center for Nanoscale Systems, Harvard University, Cambridge, Massachusetts 02138, USA\\ 
and Department of Physics, Loughborough University, Leicestershire LE11 3TU, UK}  
\date{\today}
\begin{abstract}
Spin splitting of the energy spectrum of single-layer graphene on Au/Ni(111) substrate has been recently reported. I show that eigenstates of spin-orbit coupled graphene are polarized in-plane and perpendicular to electron momentum $\bf k$; the magnitude of spin polarization $\bf S$ vanishes when $k\rightarrow0$. In a perpendicular magnetic field $\bf B$, $\bf S$ is parallel to $\bf B$, and two zero modes emerge in the Landau level spectrum. Singular $\bf B$-dependence of their magnetization suggests existence of a novel magnetic instability. They also manifest themselves in a new unconventional quantum Hall effect.

\end{abstract}


\maketitle

\narrowtext
Since the discovery of graphene with its quasi-relativistic energy spectrum of zero-gap Dirac fermions \cite{NovoOrig,Wallace47} and unconventional Quantum Hall Effect (QHE) in single- and bilayer graphene \cite{NovoQHE,Zhang05,NovoQHEbi}, this material attracts attention because of its unique electronic properties and prospective applications in nanoelectronics \cite{Carlo08,CastroNetoRMP}. Applications in spintronics depend strongly on the control of spin-orbit coupling. Spin-orbit coupling in graphene comprises intrinsic and extrinsic components \cite{DresDresSO,KaneMele1}. First one is known to be very weak in plane graphene \cite{MinSOgra,YaoSOgra}; in carbon nanotubes it is due to their curvature \cite{Ando2000,Kuemm08}. Extrinsic spin-orbit coupling originates from the interface between graphene and substrate. Recently 
Varykhalov {\it et al.} \cite{VarAuNi} 
reported spin angle-resolved photoemission spectroscopy (SARPES) data taken from single-layer graphene on Ni(111) intercalated with a Au monolayer. They reveal strong momentum-dependent in-plane spin polarization. The monolayer of Au maintains the symmetry of graphene and strongly reduces the threefold deformation of graphene (originating from its coupling to Ni substrate) and nonequivalence of $A(B)$ lattice sites resulting from it. According to Ref.~\onlinecite{VarAuNi}, the technique brings the system closer to ideal freestanding graphene than any other preparation on a solid substrate before. These results call for a systematic theory of spin polarization in freestanding graphene with extrinsic spin-orbit coupling, and in this paper I provide such a theory for free electrons and electrons in a quantizing magnetic field $\bf B$. The results are in general agreement with the data by Varykhalov {\it et al.} \cite{VarAuNi} 
and predict (i) dependence of spin polarization on the magnitude of the momentum $\bf k$ at the spin-orbit momentum scale, (ii) spin zero modes in a quantizing field indicating a novel magnetic instability, and (iii) a new unconventional QHE.

A $4\times4$ Hamiltonian of graphene with extrinsic spin-orbit coupling can be conveniently represented in terms of a Kronecker product of $2\times2$ matrices $\mbox{\boldmath$\sigma$}$ and $\bf s$ as \cite{KaneMele1}
\be
{\cal H}^0_K=\gamma(\mbox{\boldmath$\sigma$}\cdot{\bf k})+{1\over2}\lambda(\mbox{\boldmath$\sigma$}\times{\bf s}).
\label{eq1}
\ee
Here $\gamma=\hbar v$, $v\approx10^8$ cm s$^{-1}$, and $\lambda$ is a spin-orbit coupling constant; for graphene/Au/Ni(111), $\lambda\approx 13$ meV \cite{VarAuNi}. Pauli matrices of pseudospin $\mbox{\boldmath$\sigma$}=(\sigma_1,\sigma_2)$ operate on $A(B)$ lattice cites, and ${\bf s}=(s_x,s_y)$ are Pauli matrices for real spin. It is seen from first term in (\ref{eq1}) that \mbox{\boldmath$\sigma$} transforms as a momentum $\bf k$. Therefore, as was mentioned by Kane and Mele \cite{KaneMele1}, spin-orbit term can be considered as a modified Rashba term with ${\bf k}\rightarrow\mbox{\boldmath$\sigma$}$; the conventional Rashba term $({\bf k}\times{\bf s})$ is small and will be disregarded. In the same representation, $4\times4$ spin matrices are $\sigma_0{\bf s}$, with $\sigma_0$ for a unit matrix in $\mbox{\boldmath$\sigma$}$ space. The subscript $K$ in ${\cal H}_K^0$ indicates that it is acting in the vicinity of $K$ point of the Brillouin zone, FIG.~1; for $K^\prime$ point, ${\cal H}_{K^\prime}^0=\sigma_1{\cal H}_K^0\sigma_1^{-1}$.

It is convenient to change from the $4\times4$ to a $2\times2$ formalism. First, we perform an unitary transformation of the Hamiltonian, ${\cal H}_K=U{\cal H}_K^0U^{-1}$, and of spin matrices $\sigma_0{\bf s}$ with an unitary matrix
$U=[(s_0+s_z)\sigma_0+(s_0-s_z)\sigma_1]/2$, $s_0$ being a unit matrix in spin space. Then
\bea
{\cal H}_K=\left(\begin{array}{cc}
0&C\\
C^+&\lambda s_y
\end{array}\right),\,{\hat{\bf S}}=\sigma_1{\bf s},\,{\hat S}_z=\sigma_0s_z,
\label{eq2}
\eea
with $C=\gamma(s_0k_x-is_zk_y)$; here $({\hat{\bf S}},{\hat S}_z)$ are new spin matrices. Next step is eliminating lower components of eigen-bispinors $\Psi=\left(\begin{array}{c}\varphi\\
\phi\end{array}\right)$ of the equation ${\cal H}_K\Psi=\varepsilon\Psi$  
\be
\phi=(\varepsilon-\lambda s_y)^{-1}C^+\varphi={{\gamma}\over{\varepsilon^2-\lambda^2}}
\left(\begin{array}{cc}
\varepsilon k_+&-i\lambda k_-\\
i\lambda k_+&\varepsilon k_-
\end{array}\right)\varphi.
\label{eq3}
\ee
Then we arrive at a $2\times2$ Hamiltonian
\be
H(\varepsilon)={{\gamma^2}\over{\varepsilon^2-\lambda^2}}\left(\begin{array}{cc}
\varepsilon k_-k_+&-i\lambda k_-^2\\
i\lambda k_+^2&\varepsilon k_+k_-
\end{array}\right)
\label{eq4}
\ee
that depends explicitly on its eigenvalues $\varepsilon$, $H(\varepsilon)\varphi=\varepsilon\varphi$. When ${\bf B}\neq0$, operators $k_\pm=k_x\pm ik_y$ do not commute; below, ${\bf B}=B{\hat{\bf z}}$. We note that Eq.~(\ref{eq4}) is exact, and despite the fact that the problem is nonlinear in $\varepsilon$, calculation of spin polarization in this representation is more concise than in the original $4\times4$ representation.

For free carriers, ${\bf B}=0$, eigenspinors of $H(\varepsilon)$ are
\be
\varphi_\nu({\bf k})={{1}\over{\sqrt{2}}}\left(\begin{array}{c}i\nu k_-^2/k^2\\
1\end{array}\right),\,\nu=\pm1\,.
\label{eq5}
\ee
From (\ref{eq2}) and (\ref{eq4}) follows an equation $\varepsilon_\nu(\varepsilon_\nu+\lambda\nu)=\gamma^2k^2$ for the eigenvalues $\varepsilon_\nu$. Its solutions are
\be
\varepsilon_{\nu\mu}(k)={{\nu}\over{2}}\left(-\lambda+\mu\sqrt{\lambda^2+4\gamma^2k^2}\right),\,\mu=\pm1\,.
\label{eq6}
\ee
The spectrum includes two zero-gap branches and two gapped branches of the same shape. The gap equals $2|\lambda|$, and the separation between gapped and ungapped branches is $k$-independent and equals $\lambda$. The spectrum is the same as for unbiased bilayer graphene without spin-orbit coupling \cite{McCannFalko,Nilsson,Pereira}, but the region of parabolic expansion, $k\ll|\lambda|/\gamma$, is narrow because $\lambda$ is small.

It is easily seen from (\ref{eq2}) and (\ref{eq5}) that $\langle\varphi_\nu|{\hat S}_z|\varphi_\nu\rangle=0$, hence, spins are in-plane polarized. Because $\hat{\bf S}$ includes $\sigma_1$, calculation of in-plane polarization involves the lower spinor $\phi$ and is more cumbersome. Nevertheless, it is straightforward, and applying (\ref{eq3}), (\ref{eq5}) and (\ref{eq6}) results in in-plane spin polarizations ${\bf S}_{\nu\mu}({\bf k})$ for all $(\nu,\mu)$ states
\be
{\bf S}_{\nu\mu}({\bf k})={{\langle\Psi_{\nu\mu}|{\hat{\bf S}}|\Psi_{\nu\mu}\rangle}\over{\langle\Psi_{\nu\mu}|\Psi_{\nu\mu}\rangle}}={{2\mu\gamma ({\bf k}\times{\hat{\bf z}})}\over{\sqrt{\lambda^2+4\gamma^2k^2}}}\,.
\label{eq7}
\ee
Eq.~(\ref{eq7}) indicates transverse spin polarization, FIG.~1, as concluded by Varykhalov {\it et al.} \cite{VarAuNi}. 
Its magnitude is $k$-dependent. When $k\gg k_\lambda$, $k_\lambda=|\lambda|/2\gamma$ being a characteristic spin-orbit momentum, it saturates, $|{\bf S}_{\nu\mu}|\rightarrow1$. In the opposite limit, $k\ll k_\lambda$, it vanishes as $k/k_\lambda$. 

Chirality of the spinor $\varphi_\nu({\bf k})$ is defined by $\nu$, spin polarization ${\bf S}_{\nu\mu}$ by $\mu$, and the product $\mu\nu$ specifies electron and hole spectrum branches. Experiments of Ref.~\onlinecite{VarAuNi} measured the magnitude of $\lambda$, $|\lambda|\approx 13$ meV. Measuring the sign of ${\bf S}_{\nu\mu}$ would allow finding the sign of $\lambda$; indeed, it is seen from (\ref{eq6}) that $\mu/\lambda>0$ for external Fermi circles. Due to the requirements of time-inversion symmetry, spin polarization is identical near $K$ and $K^\prime$ points. It is not clear currently which of the factors  (experimental resolution, temperature, or staggered potential of Ni substrate) was the main obstacle for measuring SARPES spectra for $k\alt k_\lambda$. However, measurment of ${\bf S}_{\nu\mu}$ for $k\agt k_\lambda$ should shed additional light on the role of these factors.

Application of well developed techniques for detecting in-plane polarization ${\bf S}({\bf k})$, based on Kerr spectroscopy \cite{Kato04,CrookerSmith,Ensslin07} and spin-galvanic effect \cite{IvchGanich}, is hampered by the conductivity of metallic substrate.  Reducing the thickness of the substrate to only a few monolayers or developing insulating substrates can render them proper efficacy. 

For ${\bf B}\parallel{\hat{\bf z}}$, applying a Peierls substitution ${\bf k}=-i\nabla+e{\bf A}/\hbar c$, $\bf A$ being a vector potential, one expresses $k_\pm$ in terms of Bose operators, $k_+=(\sqrt{2}/\ell)a^+$, $k_-=(\sqrt{2}/\ell)a$, $[a,a^+]=1$; here $\ell=\sqrt{c\hbar/eB}$ is a magnetic length. Then, instead of (\ref{eq4}), one arrives at
\be
\hat{H}(\epsilon)=|\lambda|{{2\Gamma^2}\over{\epsilon^2-1}}\left(\begin{array}{cc}
\epsilon aa^+&-i\beta a^2\\
i\beta (a^+)^2&\epsilon a^+a
\end{array}\right)\,,
\label{eq8}
\ee
where $\beta=\lambda/|\lambda|$, $\epsilon=\varepsilon/|\lambda|$, and $\Gamma=\gamma/\ell|\lambda|$. The Hamiltonian ${\hat H}(\epsilon)$ depends explicitly on its eigenvalues $\epsilon$. From here on, energy $\epsilon$ is measured in the units of $|\lambda|$.

Solution of the corresponding eigenspinor problem can be found in terms of oscillator eigenfunctions $\psi_n$ \cite{Lutt56}
\bea
\varphi_n&=&\left(\begin{array}{c}c_1\psi_{n-2}\\c_2\psi_n\end{array}\right),\phi_n=\left(\begin{array}{c}c_3\\c_4\end{array}\right)\psi_{n-1},
\nonumber\\
\left(\begin{array}{c}c_3\\c_4\end{array}\right)&=&{{\Gamma\sqrt{2}}\over{\epsilon^2-1}}
\left(\begin{array}{c}\epsilon\sqrt{n-1}~c_1-i\beta\sqrt{n}~c_2\\
i\beta\sqrt{n-1}~c_1+\epsilon\sqrt{n}~c_2\end{array}\right),
\label{eq9}
\eea
for $n\geq2$. The coefficients $c_{1,2}=c_{1,2}(n)$, normalized as $|c_1|^2+|c_2|^2=1$, read as
\be
{{c_1}\over{c_2}}={{i\beta\epsilon(1+2n\Gamma^2-\epsilon^2)}\over{2\sqrt{n(n-1)}\Gamma^2}},\,
c_2={{\sqrt{n}}\over{2n(1+\Gamma^2)-\epsilon^2}}\,.
\label{eq10}
\ee
Eigenvalues obey the equation 
\be
\epsilon^4-[1+2\Gamma^2(2n-1)]\epsilon^2+4n(n-1)\Gamma^4=0
\label{eq10a}
\ee
and are 
\be
\left(\epsilon_n^\pm\right)^2={1\over2}\left[1+2(2n-1)\Gamma^2\pm\sqrt{1+4(2n-1)\Gamma^2+4\Gamma^4}\right].
\label{eq11}
\ee
Eq.~(\ref{eq11}) coincides with the expression for bilayer graphene in absence of spin-orbit coupling \cite{Pereira,Nakamura08,Henrik08}.

In addition to the solutions with $n\geq2$, there are solutions with $n\leq1$ that are of special interest because of their peculiar spin properties. For $n=0$, there is a single solution $c_1(0)=0$, $c_2(0)=1$, $\phi_0=0$, with $\epsilon_0=0$. For $n=1$, there are three solutions with $c_1(1)=0$ and $c_2(1)=1$, they differ by their $\phi_1$ spinors. For one of them eigenvalue vanishes, $\epsilon_1^0=0$, and components of the spinor $\phi_1^0$ are $c_3=i\beta\Gamma\sqrt{2}$, $c_4=0$. Two nonvanishing eigenvalues are $\epsilon_1=\pm\sqrt{1+2\Gamma^2}$; the components of the corresponding spinors $\phi_1$ are $c_3=-i\beta/\Gamma\sqrt{2}$, $c_4=\epsilon_1/\Gamma\sqrt{2}$. These expressions can be also found from (\ref{eq9}) and from (\ref{eq11}) [with the upper sign in (\ref{eq11})] by plugging $n=1$. Hence, there exist two zero modes, $\epsilon_0=0$ and $\epsilon_1^0=0$, and spin-orbit coupling of $(\mbox{\boldmath$\sigma$}\times{\bf s})$-type preserves two-fold degeneracy of $\epsilon=0$ state typical of single-layer graphene \cite{NovoQHE,Zhang05,GusShar05}, but changes its nature: Degeneracy is dynamical rather than Kramers (small Zeeman splitting \cite{Zhang06} is disregarded).

It follows from (\ref{eq9}) that in-plane spin polarization $\bf S$ vanishes in all eigenstates. Indeed, because of the factor $\sigma_1$ in ${\hat{\bf S}}=\sigma_1{\bf s}$, it mixes different components of the bispinor $\Psi$, $\varphi$ and $\phi$, and mean value of their product is proportional to scalar products of the oscillator functions $\psi_m$ with quantum numbers that never coincide. Therefore, ${\bf S}_n=0$ for all $n$. This result is expected because ${\bf S}_{\nu\mu}({\bf k})$ of (\ref{eq7}) vanishes after averaging over the direction of $\bf k$, and follows from axial symmetry of the problem.

Longitudinal polarization $S_z$, after eliminating $\phi$ component of $\Psi$, can be expressed in terms of its $\varphi$ component 
\be
S_z=\left\langle\varphi\left|s_z+{{\Gamma^2}\over{\epsilon^2-1}}[s_0+(aa^++a^+a)s_z]\right|\varphi\right\rangle
\bigg/\langle\Psi|\Psi\rangle,
\label{eq12}
\ee
where $\epsilon$ is energy of the eigenstate $\varphi$. The normalization factor equals
\bea
\langle\Psi|\Psi\rangle&=&1+{{\Gamma^2}\over{(\epsilon^2-1)^2}}\bigg\langle\varphi\bigg|2i\beta\left[(a^+)^2s_--a^2s_+\right]\nonumber\\
&+&(\epsilon^2+1)\left[(aa^++a^+a)s_0+s_z\right]\bigg|\varphi\bigg\rangle\,.
\label{eq13}
\eea
Explicit expressions for (\ref{eq12}) and (\ref{eq13}) follow from (\ref{eq9}) and (\ref{eq10}). Their original form is cumbersome but greatly simplifies after higher powers of $\epsilon^2$ are eliminated by employing (\ref{eq10a}), and $\beta^2=1$ is applied. The final equation, when expressed in terms of $\epsilon_n^\pm$, is rather concise 
\be
(S_z)_n^\pm={{2\Gamma^2(2n\Gamma^2-\epsilon^2)}\over{\epsilon^2(1-2\Gamma^2)+2n\Gamma^2(1+2\Gamma^2)}}\,,\,\epsilon=\epsilon_n^\pm.
\label{eq14}
\ee
This equation, together with (\ref{eq11}), provides exact expressions for $B$-dependence of spin polarization for all states with $n\geq1$ and $\epsilon\neq0$. Because $\epsilon$ appears in (\ref{eq14}) only as $\epsilon^2$, polarization $S_z$ is charge symmetrical: It coincides for electron and hole states with the same $n$ and $|\epsilon^\pm_n|$. 

In the weak field limit, $n\Gamma^2\ll1,\epsilon^2$, eigenvalues are $\epsilon_n^+\approx1$ and $\epsilon_n^-\approx2\sqrt{n(n-1)}\Gamma^2$, and $S_z$ is $n$-independent and proportional to $B$, $(S_z^\pm)_n^\pm\approx\mp2\Gamma^2$ ($n>1$ for $\epsilon^-_n$ states). Because of the spectrum degeneracy at $k=0$ point, the sequence $\epsilon^-_n$ is nonequidistant despite the parabolicity of the spectrum \cite{Lutt56}. In the strong field limit, $\Gamma^2\rightarrow\infty$, eigenvalues are $\epsilon_n^-\approx\Gamma\sqrt{2n}$ and $\epsilon_n^+\approx\Gamma\sqrt{2(n-1)}$, and spin magnetization saturates, $(S_z)^\pm_n\approx\mp1$. In this limit, two ladders nearly overlap and are split by $\Delta\epsilon_n\approx\sqrt{n/2}/(2\Gamma)$; the splitting increases with $n$ but for $\Gamma^2\gg n$ is small compared with the level separation $\Gamma/\sqrt{2n}$ inside each ladder. For $\lambda\approx 13$ meV, the field separating these two limit cases, found from the condition $\Gamma=1$, equals $B_{\rm cr}=c\lambda^2/(e\hbar v^2)\approx 0.3$ T.

In both limits, spin magnetization has opposite sign for $\epsilon^+$ and $\epsilon^-$ ladders, hence, magnetization oscillates when Landau levels cross the Fermi level. These de Haas-van Alphen type oscillations can be detected by Kerr effect spectroscopy \cite{AwKikk} even for ferromagnetic substrates.

Semiclassical regime is achieved for $\Gamma\ll1$ and $n\gg1$ with $n\Gamma^2=\kappa^2/2=$ const; for $\kappa=\gamma k/|\lambda|$, one recoveres (\ref{eq6}) from (\ref{eq11}). Keeping $\kappa$ = const, one finds for electron branches, $\epsilon_n^\pm>0$, in the first order in $1/n$,
\bea
\epsilon_n^\pm&\approx&{1\over2}\left(\sqrt{1+4\kappa^2}\pm1\right)-{{\kappa^2}\over{2n\sqrt{1+4\kappa^2}}}\,,\nonumber\\
(S_z)_n^\pm&\approx&~\mp{{\kappa^2}\over{n\sqrt{1+4\kappa^2}}}=\mp{{2\Gamma^2}\over{\sqrt{1+4(\gamma k/\lambda)^2}}}\,.
\label{eq15}
\eea
Therefore, in the leading order of the expansion, level splitting $\Delta\epsilon_n\approx1$ remains $n$-independent, while spin polarization $S_z$ has opposite sign for two spectrum branches and decreases with $n$ (or, for $\Gamma$= const, with the electron momentum $k$). The oscillatory dependence of the total spin magnetization on $B$ can be detected through Kerr spectra as discussed in the previous paragraph.

Therefore, all quantum states with $\epsilon_n^\pm\neq0$ are nondegenerate, and only two states $\epsilon_0=\epsilon_1^0=0$ are degenerate. These two states differ strongly in their spin magnetization. For the $n=0$ state, polarization $(S_z)_0=-1$. It does not depend on $B$, but changes abruptly when $\bf B$ changes sign. For the $n=1$ state with $\epsilon_1^0=0$, polarization equals $(S_z)_1=-(1-2\Gamma^2)/(1+2\Gamma^2)$; it tends to $-1$ for $B\rightarrow0$, changes sign at $\Gamma^2=1/2$, and saturates to $+1$ for $B\rightarrow\infty$. In this limit, contributions of two $\epsilon=0$ states cancel. Such a behavior is unique because it suggests that spin-orbit coupled graphene with filled $\epsilon=0$ states is unstable to ferromagnetic ordering in $\hat{\bf z}$ direction in weak fields ${\bf B}\parallel{\hat{\bf z}}$, while with increasing $B$ the magnetization gradually vanishes. This magnetization differs drastically from the edge-state magnetization of graphene zigzag nanoribbons proposed by Fujita {\it et al.} \cite{Fujita96} because it originates from spin-orbit coupling rather than from exchange interaction. It also bears no similarity with the Dzyaloshinskii-Moriya weak ferromagnetism \cite{Dzia57,Moriya60} because it is dynamical rather than symmetry conditioned, and develops in a paramagnet without any magnetic structure. Effect of electron-electron interaction on this peculiar state needs a special investigation. 

For comparison with experimental data, one needs to add magnetization $S_z$ of the electrons in $K$ and $K^\prime$ valleys. Because ${\cal H}^0_{K^\prime}$ is related to ${\cal H}^0_{K}$ by a $\sigma_1$ canonical transformation leaving the operator ${\hat{S}}_z$ of (\ref{eq2}) unchanged, magnetization has the same magnitude and sign in both valleys. This can be also inferred from the fact that $K$ and $K^\prime$ valleys are related by a $\pi/3$ rotation about the $z$ axis that does not change $S_z$. Therefore, the total $S_z$ equals the magnetization of $K$ valley multiplied by a factor of 2.

Novoselov {\it et al.} \cite{NovoQHEbi} compared the conventional QHE \cite{Klitzing80} and two types of unconventional QHEs typical of single-layer \cite{NovoQHE,Zhang05} and bilayer graphene, the material notorious for its exotic QHE properties \cite{Hald04}. Spin-orbit coupled single-layer graphene introduces one more type of unconventional QHE. Due to the four-fold degeneracy of zero modes [$\epsilon_0=\epsilon_1^0=0$ degeneracy times factor of 2 from isospin, $K(K^\prime)$ valleys] the step in $\sigma_{xy}$ at $B=0$ equals $4e^2/h$, as in single-layer graphene without spin-orbit coupling. However, because spin degeneracy is lifted in each valley by spin-orbit interaction, and only isospin degeneracy persists, all $B\neq0$ steps are of $2e^2/h$. Therefore, the ratio of the magnitudes of $B=0$ and $B\neq0$ steps equals 2, as in bilayer graphene without spin-orbit coupling. Depending on the magnitude of $\lambda$, these $2e^2/h$ steps can appear in pairs, as resolved $4e^2/h$ steps, similarly to the resolution of two spin components of the traditional QHE. Spin-orbit coupled graphene on an isolating substrate would become the optimal object for observing this new QHE and for spin manipulation, a challenging task for semiconductor spintronics.

Apparently, among the perturbations that lower the symmetry of the Hamiltonian the largest is the staggering sublattice potential of the substrate violating the equivalence of $A(B)$ lattice cites; manifestation of the broken six-fold symmetry was reported in Ref.~\onlinecite{VarAuNi}. A staggered potential can be described by a term ${\cal H}_{\rm st}=u\sigma_z\tau_z$, where $\tau_z$ is a Pauli matrix in the isospin space \cite{KaneMele1}. This term creates a gap in the spectrum and lifts the symmetry of $K(K^\prime)$ valleys. Adding ${\cal H}_{\rm st}$ to ${\cal H}^0_{K(K^\prime)}$ does not change wave functions of $n=0$ modes but changes their energies to $\varepsilon=\pm u$ (in dimensional units). For the $n=1$ soft mode, eigenvectors also change; now $c_1=0$ but $c_2,c_3,c_4\neq0$. Energy spectrum can be found from a cubic equation that for $|u/\lambda|\ll1$ defines soft modes $\varepsilon_1\approx\pm u(1-2\Gamma^2)/(1+2\Gamma^2)$. As a result, the $4e^2/h$ step in $\sigma_{xy}$ splits into a plateau near $B=0$ and two $e^2/h$ steps on both sides of it. Lifting the $K(K^\prime)$ degeneracy also splits all $2e^2/h$ steps, making the QHE of spin-orbit coupled graphene similar to the traditional QHE with a resolved Zeeman splitting. We notice that $\lambda=13$ meV corresponds to a field $B=120$ T for a Land\'{e} factor $g=2$.

In conclusion, a theory of the energy spectrum and spin polarization in single-layer graphene a subject to a substrate-induced spin-orbit coupling is presented.  Energy spectrum consists of two zero-gap bands and two gapped bands reminding the spectrum of bilayer graphene without spin-orbit coupling. However, all states are in-plane spin-polarized perpendicular to the momentum $\bf k$.  This polarization saturates at large $k$ and vanishes at the scale of spin-orbit energy when $k\rightarrow0$. In a perpendicular magnetic field, two zero modes develop in each of $K(K^\prime)$ valleys. These modes show a peculiar magnetic behavior suggesting a possibility of a perpendicular-to-plane spin-orbit conditioned magnetism, and produce a new unconventional quantum Hall effect.

I am grateful to F. Kuemmeth, C. M. Marcus, and J. R. Williams for stimulating discussions.

FIG.~1. Spin polarization of energy spectrum of spin-orbit coupled single-layer graphene with the dispersion law of Eq.~(\ref{eq6}) and spin-orbit coupling constant $\lambda>0$; energy $\varepsilon>\lambda$. Left column - $K^\prime$ valley, right column - $K$ valley. Upper row - electrons, $\mu\nu=+1$; lower row - holes, $\mu\nu=-1$. External circles - $\mu=+1$; internal circles - $\mu=-1$. For $\lambda<0$, quantum numbers, $\mu$ and $\nu$, and spin polarizations change their signs for all branches. Brillouin zone of graphene is also shown.

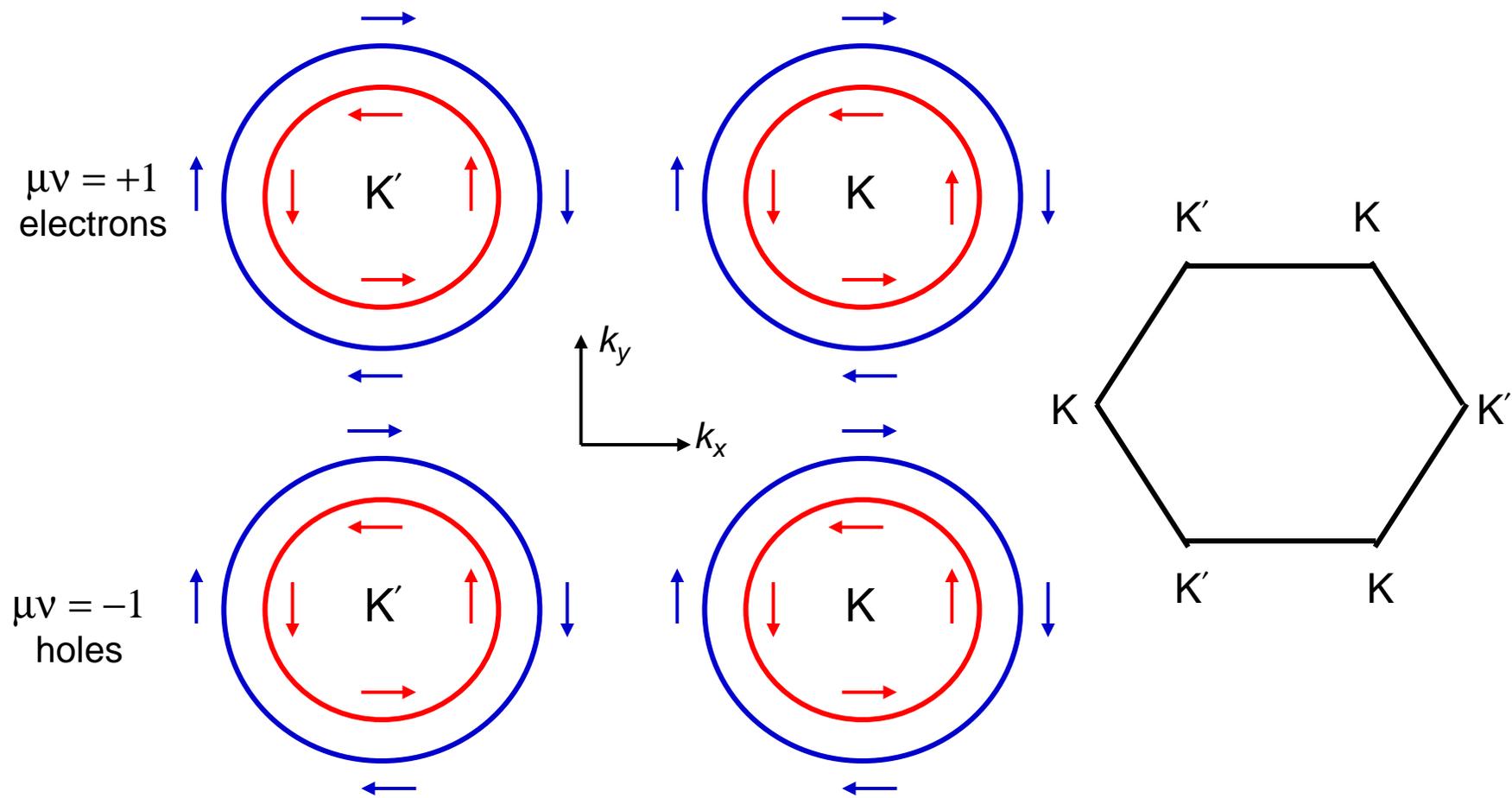
\end{document}